\documentclass[fleqn,usenatbib]{mnras}
\usepackage{graphicx}
\usepackage{txfonts}
\usepackage{geometry}                		
\usepackage{graphicx}
\usepackage{amssymb}
\usepackage{multicol}       
\usepackage{bm}		
\usepackage{pdflscape}
\usepackage{ulem}
\usepackage{hyperref}
\usepackage{CJKutf8}
\usepackage{rotating}
\usepackage{tabularx}
\usepackage{ragged2e}
\usepackage{lipsum}
\usepackage{booktabs} 
\usepackage{ae,aecompl}
\usepackage{titlesec}
\usepackage{float}
\usepackage{chngcntr}
\usepackage{bm}		
\usepackage{pdflscape}	
\usepackage{rotating}
\usepackage{tabularx}
\usepackage{ragged2e}
\usepackage{lipsum}
\usepackage{booktabs} 
\usepackage{ae,aecompl}
\usepackage{titlesec}
\usepackage{hyperref}
\usepackage{float}
\usepackage{chngcntr}

\newcommand{\msun}{M_\odot}

\newcommand{\mcluster}{M_{\rm cluster}}

\newcommand{\nbody}{\texttt{NBODY6}}
\newcommand{\nbodyplus}{\texttt{NBODY6++}}
\newcommand{\nb}{\texttt{NBODY6++GPU}}

\newcommand{\rebound}{\texttt{REBOUND}}

\newcommand{\sph}{\texttt{SPH}}
\newcommand{\gasph}{\texttt{GaSPH}}

\newcommand{\snipes}{\texttt{SnIPES}}
\newcommand{\nstars}{N_{\rm s}}

\newcommand{\crossingtime}{t_{\rm cr}}
\newcommand{\relaxationtime}{t_{\rm rh}}

\usepackage{xcolor}  
\usepackage{hyperref}
\usepackage{blindtext} 
\usepackage{comment}

\usepackage{academicons}
\usepackage{xcolor}
\usepackage{orcidlink}
\newcommand{\orcid}[1]{\href{https://orcid.org/#1}{\textcolor[HTML]{A6CE39}{\aiOrcid}}}
\title[Disks and planets in clusters]{The dynamical evolution of protoplanetary disks and planets in dense star clusters}

\author[Flammini Dotti et al.]{Francesco Flammini Dotti$^{1,2,3}$\thanks{Contact e-mail: \href{mailto:fmfd@uni-heidelberg.de}{fmfd@uni-heidelberg.de}}\orcidlink{0000-0002-8881-3078}
          ,
          R. Capuzzo-Dolcetta$^{4}$\orcidlink{0000-0002-6871-9519}
          \and
          M. B. N. Kouwenhoven$^{2}$\orcidlink{0000-0002-1805-0570} \\
\\
$^1$Astronomisches Rechen-Institut, Zentrum f\"ur Astronomie, University of Heidelberg, \\ M\"onchhofstrasse 12--14, 69120, Heidelberg, Germany \\
$^2$Department of Physics, Xi'an Jiaotong-Liverpool University, 111 Ren'ai Rd., \\
Suzhou Dushu Lake Science and Education Innovation District, Suzhou Industrial Park, Suzhou 215123, P.R. China\\
$3$Department of Mathematical Sciences, University of Liverpool, Liverpool L69 3BX, UK\\
$4$Dipartimento di Fisica, Sapienza, Universita di Roma, P.le Aldo Moro, 5, 00185 - Rome, Italy\\
}

\date{Received ---; accepted ---}

\begin{document} 

\label{firstpage}
\pagerange{\pageref{firstpage}--\pageref{lastpage}}
\maketitle

\begin{abstract}

Most stars are born in dense stellar environments where the formation and early evolution of planetary systems may be significantly perturbed by encounters with neighbouring stars. \\
To investigate on the fate of circumstellar gas disks and planets around young stars dense stellar environments, we numerically evolve star-disk-planet systems. We use the $N$-body codes \nb{} and \snipes{} for the dynamical evolution of the stellar population, and the \sph{}-based code \gasph{} for the dynamical evolution of protoplanetary disks.\\
The secular evolution of a planetary system in a cluster differs from that of a field star. Most stellar encounters are tidal, adiabatic and nearly-parabolic. The parameters that characterize the impact of an encounter include the orientation of the protoplanetary disk and planet relative to the orbit of the encountering star, and the orbital phase and the semi-major axis of the planet. We investigate this dependence for close encounters ($r_p/a\leq 100$, where $r_p$ is the periastron distance of the encountering star and $a$ is the semi-major axis of the planet). We also investigate distant perturbers ($r_p/a\gg 100$), which have a moderate effect on the dynamical evolution of the planet and the protoplanetary disk. We find that the evolution of protoplanetary disks in star clusters differs significantly from that of isolated systems. When interpreting the outcome of the planet formation process, it is thus important to consider their birth environments.
\end{abstract}


\section{Introduction}

A significant fraction of stars in the Milky Way is thought to host one or more planetary companions \citep[e.g.,][]{mayo2018, thompson2018}, and even binary star systems can host exoplanets \citep{gould2014}. Over 5380 exoplanets have now been identified in 3974 extra-solar planetary systems, among which 857 are multi-planetary systems\footnote{http://exoplanet.eu, accessed on 17 May 2023}. Most stars form in clustered environments \citep[e.g.,][]{lada2003}. The majority of these embedded star-forming regions dissolve within $50$~Myr \citep[e.g.,][]{leis1989,grijs2008, grijs2009}, while the remainder evolves into open clusters. Observational evidence also suggests that our Solar system may have formed in a clustered environment \citep[e.g.,][]{adams2010, pfalzner2013, pz2018}. The planet formation process and the early dynamical evolution of star-forming regions occur at comparable timescales. It is therefore of interest to model both processes simultaneously. 

During the early evolution of protoplanetary systems, gravitational interactions with neighbouring stars can affect the evolution of protoplanetary disks and young planetary systems \citep[e.g.,][]{thies2005, olczek2012, pz2016, vinncke2018}. These close encounters may leave imprints on planetary systems that later become part of the much older population in the Galactic neighbourhood. Stellar encounters may perturb or even disrupt protoplanetary disks and planetary systems. This mechanism results in free-floating planets in star clusters. When they have sufficiently high speeds, these free-floating planets can rapidly escape from their parental cluster \citep{wang2015a, kouwenhoven2020}. The free-floating planets may also migrate to the outskirts of the star cluster, to be eventually stripped off by the Galactic tidal field or recaptured by other stars \citep[e.g.,][]{perets2012}.  

Although substantial progress has been made in recent years,  modeling the dynamical evolution of young planetary systems in dense stellar environments remains computationally complex. Numerical challenges arise from the large dynamical ranges in the length scale, the time scale, and the mass range that have to be implemented in the code. A second difficulty is the inclusion of gas in the model of the cluster and the planetary systems. A fully self-consistent simulation of young, gas-rich star clusters with planetary systems remains challenging. Different approaches have been taken to partially overcome these challenges: (i) modeling of isolated planetary systems; (ii) scattering experiment for modeling the evolution of multi-planet systems; (iii) modeling of single-planet systems in star-cluster environments; and (iv) separately modeling the star clusters and planetary systems, under the assumption that the planetary dynamics do not affect the the stellar components. Similar approaches can also be used to model the evolution of circumstellar gas disks in star clusters, as demonstrated in this study.
\cite{spurzem2009} present a comprehensive study of planetary system evolution in star clusters. They find that numerical results obtained with the direct $N$-body code \nbodyplus \citep{spurzem1999} and with a hybrid Monte Carlo code \citep{spurzemgiersz1996, gierszspurzem2000, gierszspurzem2003} are consistent with theoretical estimates. 
\cite{zheng2015} used $\nbody$ to build the evolution of single-planet systems in multi-mass open star clusters, and  derived analytical prescriptions for the retention rate of planetary companions and free-floating planets as a function of initial semi-major axis and cluster properties. 
\cite{fujii2019} followed a more comprehensive approach for developing an analytical prescription of the escape probability. Their study focus on the Pleiades, Hyades and Praesepe clusters, which are thought to have formed in highly-substructured star-forming regions \citep[e.g.,][]{fujii2012,sabbi2012,fujii2015}. The study focused on single-planet systems orbiting Solar-like stars. The escape probability dependence on the semi-major axis, $a_p$, follows the distribution $p_{esc} \propto a_p^{-0.76}$, which is consistent with that of \cite{caietal2017}. 
\cite{puu} model two-planet systems in star clusters using \rebound{} \citep{rein2012}, and compare their results to hybrid secular equations. Their approach provided insight into the origin of super-Earths and sub-Neptunes, and  the Kepler-11 system in particular. They explain why multiple transiting planets appear to be dynamically \lq colder \rq than those with a single transiting planet.
\cite{caietal2017} modeled open star clusters with Solar-like stars that host five equal-mass planets separated by $10-100$ mutual Hill radii. Most host stars retain their planets, although stellar encounters and planet-planet interactions trigger perturbations in eccentricity and inclination that occasionally lead to a decay of the system. \cite{caietal2018} and \cite{caietal2019} analysed how the signatures of the star cluster affects the observed characteristics of exoplanet systems in the Galactic field. 
\cite{flaetal2019} studied the impact of stellar encounters on the evolution of planetary systems similar to our Solar system. Their study shows that planet-planet scattering is a important consequence of perturbation on previously stable planetary systems, and that the stability of the system depends on the orbital architecture and the planetary mass spectrum. Similarly, \cite{wukai2023} found that for planetary systems in star clusters, planets can affect the dynamical evolution debris particles far beyond their Hill radii.
The approaches used for modeling the evolution of planetary systems in star clusters can also be used to study the evolution of protoplanetary systems in such environments. protoplanetary systems are associated with gas accretion onto proto-stars \citep{armitage2019}. Gas is accreted onto the star and partially ejected from its poles due to the magnetic field. The gas is eventually distributed in a disk-like shape around the star, and a protoplanetary disk is formed, composed of gas and dust. Turbulence arises from hydro-magnetic instabilities, which eventually leads to dust agglomeration. Pebbles and planetesimals start to form, leading to the formation of the protoplanetary cores and ultimately planets \citep{papa2005, papa2007}. 
The early evolution of protoplanetary systems is affected by the neighbouring stellar population. A good example of this is the Neptune-like planet in a binary system near the Hyades cluster \citep{ciardi2018}. Current theories do not predict this formation scenario. Therefore, these types of planetary systems raised the need to study how the imprint of an close encounter would eventually lead to a different evolutionary scenario.\\
A number of disks in dense star clusters have been detected \citep[e.g.,][]{hernandez2010, mann2015}. HARPS-N \citep{pepe2000} observed the first multi-planet system in a young massive cluster \citep[M44; see][]{malavolta2016}. 
Low-mass disks, such as those analysed in isolated cases \citep[e.g.,][]{antonyuk2015} have a negligible gravitational feedback on the hosting stars. However, these protoplanetary disks can be substantially truncated in a timescale longer than the crossing time \citep{rosotti2014}.\\
Smoothed Particle Hydrodynamics (SPH) codes can be used to model the dynamical evolution of protoplanetary disks. These codes are fully Lagrangian, and are particularly suited for non-symmetric systems and for dealing with self-gravity  \citep[for recent reviews see, e.g.,][]{monaghan2005, spri10,price2007, pri11}. A recent application of SPH to the study of the feedback of a protoplanetary disk around a \textit{target} star by a close passage of a \textit{bullet} star is found in \cite{cat2020}.\\
In this paper, we investigate the evolution of protoplanetary systems in dense stellar environments. We combine $N$-body simulations for the stellar dynamics in a cluster with SPH treatment for the gaseous disk dynamics around a star. We aim to understand how planetary systems dynamically interact with both star cluster and the evolving protoplanetary disk. This paper is organized as follows. In Section~\ref{methundmod}  we present our methodology. In Section~\ref{results}  we discuss our various models and their results. Finally, we summarize and discuss our conclusions in Section~\ref{conclusions}.

\section{Methodology and initial conditions} \label{methundmod}

\subsection{Initial conditions - Star cluster}

\begin{table}
    \caption{
    (\textit{Top table}) Initial conditions for the star cluster model: the model ID (column~1, using the syntax C-$Q$, where $Q$ is the virial ratio), the initial number of stars (column~2), the initial total star cluster mass (column~3), the initial crossing time and the initial half-mass relaxation time (columns~4 and~5).  \\
    (\textit{Central table}) Initial conditions for the three encounter classes, based on different distances between the bullet star and the host star) which the planetary system is subjected to: the disk case ID (column~1, using the syntax M-\#, where the \# stands for the model number), the host star mass (column~2), the star cluster age at the time of the encounter (column~3), the host star position in the star cluster (column~4), and the encountering star mass (column 5). \\
    (\textit{Bottom table}) Main characteristics of the protoplanetary disk, in column order: the number of gas particles, the viscosity parameter, the internal and external cut-offs of the disk, the disk mass, and the planet mass. We will use model M1 as our reference model for the following sections, unless specified otherwise.}
    \begin{tabular}{lrccl}
    \hline
    Model ID & $\nstars$ & $\mcluster$  & $\crossingtime$ & $\relaxationtime$   \\
    &     & $\msun$ & Myr & Myr   \\
    \hline
    C05 & 10\,000 & $5.87\times10^3 $ & 0.18 & 26.59  \\
    \vspace{0.5cm}
    \\
    \hline
    ID & $M_{*}$ & Age & $r_{\rm hs}$ & $M_{\rm 1,enc}$ \\
    &   $\msun$ & Myr & pc & $\msun$ \\
    \hline
    M1 & 0.99 & 23.66 &  0.20 & 0.69 
     \\
    M2 & 0.97 & 24.01 & 0.02 & 0.63 
    \\
    M3 & 0.99 & 20.88 & 0.34 & 1.03 
    \\
    \vspace{0.5cm}
    \\
    \hline
    $N_{d}$ & $\alpha_{\sph}$ & $r_{in},r_{out}$ &  $M_{disk}$ & $M_p$ \\
    &   & AU & $\msun$ & $M_{j}$ \\
    \hline
    $50\,000$ & $ 0.1$ & $10,100$ &    $10^{-3}$ & 0.05 \\
    \end{tabular}
    \label{Table1}
\end{table}

We model a star cluster containing 10\,000 stars. We adopt the \citep{plummer1911} density profile in virial equilibrium, with a virial radius of 1~pc. Stellar masses are drawn from the \cite{kroupa2001} initial mass function (IMF), in the mass range $0.1-25\,\msun$, and we adopt a Solar metallicity. For simplicity, our models do not include primordial binary systems. The star cluster is evolved in an external tidal field (the Standard Solar tidal field). The initial conditions of the star cluster are summarized in the top table of Table~\ref{Table1}. We evolve the models for 50~Myr using \nb{} \citep{aarseth1999, spurzem1999, kamlah2022b}.

\subsection{Initial conditions - protoplanetary disk and planet}

\begin{figure}
    \includegraphics[width=0.5\textwidth,height=!]{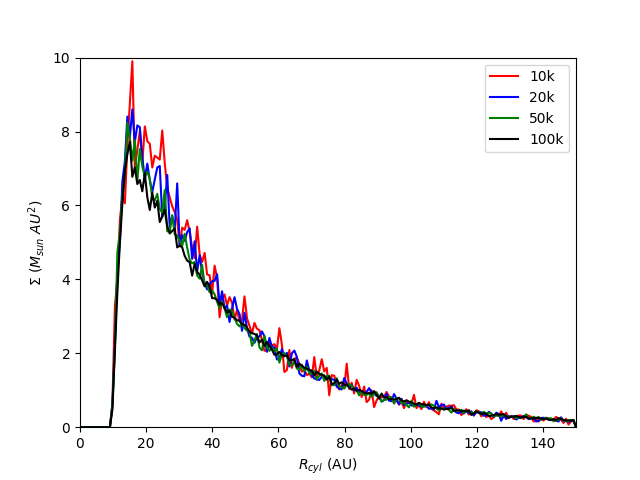} 
    \caption{Surface density distributions of the cut circum-stellar disk for different particle numbers at 30\,000~yr.
}
    \label{Figure1}
\end{figure}

The circumstellar gas disk is modeled with $N_{\rm sph}=50\,000$ gas SPH particles, and has a mass $M_{\rm disk}=10^{-3}~\msun$. This particle number has been shown to be sufficient \citep[see][]{pinto2019,cat2020} to reach good convergence and stability of a reasonable disk model of the type discussed hereafter. Figure~\ref{Figure1} indicates, indeed, how the surface distribution sample with a number of SPH particles from 50\,000 well represents the proto-planetary disk under study.
The disk revolves around a $1\,\msun$ star in all models, and the star constitutes the centre of the reference system. The density follows a classical flared disk distribution model, with the following azimuthally-symmetric distribution:
\begin{equation}
  \rho (R,z) = \frac{\Sigma (R)}{H} \exp{\left(-\frac{z^2}{2 H^2}\right)} \quad , 
\end{equation}\label{eq_diskrho}
where $\Sigma(R)$ is the cylindrical radial surface density profile, and $H$ is the vertical scale height, locally dependent on the cylindrical coordinate $R$. The disk is modelled with a radial density profile $\Sigma \sim R^{-p}$ with $p=-3/2$, according to the classical \cite{hayashi1981} scheme. The initial gas surface density profile is distributed between an inner cut-off radius $r_{\rm in}$ and an outer cut-off radius of $r_{\rm out}$, such that $r_{\rm out}/r_{\rm in}=10$. Further details on the initialization of circumstellar disks are discussed in \cite{pinto2019}.\\
To avoid excessively short time-steps in the regions close to the stars, a suitable computational \lq{}sink radius\lq{} is set up. All the gas particles that approach a star within the sink radius, provided that they are gravitationally bound, are accreted onto the object, and are subsequently excluded from the integration. \\
The inner cut-off radius corresponds to the sink radius of the central star. The role of this quantity on the results is of minor relevance, as we are primarily focused on the global evolution of the disk and the planet, which depend more on the large-scale structure of the disk (its outward extension and former stability). The initial outer cut-off radius is motivated by our region of interest. \\
The protoplanetary disk is initialized such that gas particles orbit the host star in roughly Keplerian orbits. The dynamical evolution of the star cluster and the consequent close encounters with neighbouring stars, may change the evolution of the gas disk over time, in addition to the internal processes that evolve the disk.\\
We study the evolution of systems containing a single, Neptune-mass planet in orbit around the host star. The planet's mass is selected to obtain a system with a well-defined mass hierarchy between the star, the proto-planetary disk and the planet. The orbital properties of the planetary system are discussed in Section~\ref{M1section}.

\subsection{Numerical method}\label{methodology}

We perform the simulations of star clusters, planets and protoplanetary disks, by combining the $N$-body code \nb{} \citep{kamlah2022b}, the new $N$-body code \snipes{}, and the SPH code \gasph{}. The codes and the numerical approach are discussed below.

\subsubsection{Star cluster simulations}

We first model the star cluster environment using \nb{} \citep{kamlah2022b}. \nb{} is the most recent update of the original \nbody{} \citep{aarseth1999} and \nbodyplus \citep{spurzem1999}. The kinematic data of the star cluster members are then stored for subsequent high-resolution modeling the trajectories of neighbouring stars during their encounters with a planetary system.

\subsubsection{Decision-making on the host star and its neighbours}

We select a host star through a encounter strength estimation via the $k$ parameter; see Section~\ref{sec:kpar} for details. The $k$ parameter give us a range of encounters which we can choose from, while also determining the closest encounters. The selection is taken from the most effective encounter in the case of M2 and M3. We filter and select the encounter which tends to be (or is near) the impulsive, non-adiabatic and parabolic regions \citep{flaetal2019}. For M1 we use the same approach, but we also take into account which of the strongest encounters is the most probable. This is to ensure to use the statistically more probable short-distance strong encounter.  
After selecting a host star (the star with a planet and protoplanetary disk), we identify its stellar neighbour stars within a sphere of radius $r_s \sim$ 40\,000~AU at the time of closest approach.

\subsubsection{Integrating the host star and neighbour sphere: \snipes}

When integrating the trajectories, we include the neighbour stars identified above. Perturbations from more distant stars are ignored. Since the kinematic data from \nb{} are not stored at a sufficiently high temporal resolution as required by the SPH code, we have therefore developed an $N$-body integrator \snipes{} ({\it Stars 'Nd Inner Planets Evolution Solver}),  which integrates the orbits of the neighbouring stars. \snipes{} uses REBOUND \citep{rein2012} to handle particle integration, using the IAS15 high order integrator \citep{rein2015}. The procedure shares some similarities with the code LonelyPlanets \citep[e.g.,][]{caietal2015, caietal2016, caietal2017, caietal2018, caietal2019, flaetal2019}, but it has been modified to focus on the stellar neighbours (of the order of a hundred for our code, unlike in LonelyPlanets, where a maximum of ten neighbours are integrated). We plan to release an article on this code in the next future.\\
We then integrate all stars within the sphere for 1~Myr, starting 500~kyr before the closest approach, and ending 500~kyr after the closest approach. We integrate and store the trajectories at a higher time resolution, so that the time intervals are shorter than the dissipation time of the disk \citep{pinto2019}.

\subsubsection{The SPH simulation}

After having obtained the stellar trajectories, we start the second phase of our investigation. We consider the approaching stars, along with the neighbour stars which affect both the host and encountering star. We introduce $t_{fb}$ as the time at which the host star and encountering star reach their minimum mutual distance. We also introduce the quantity $\delta t_{fb}$, which represents a suitable interval of time before the close encounter, such that gravitational attraction between the two stars is negligible compared to their kinetic energies. We focus on the status of the $N$-body system at a slightly earlier time $t_{PRE-fb} = t_{fb} - \delta t_{fb}$. We consider the $N$-body simulation at the time $t_{PRE-fb}$ by restricting our study to an ensemble of stars within a range from the center of mass of the two encountering objects. 
We register positions, velocities and masses of those objects. This output will set the initial conditions of the \sph{} simulations.\\
The hydro-dynamical evolution of protoplanetary disk interacting with a small ensemble of stars, using the \gasph ~code, is described in \cite{pinto2019}. They use a \sph{} scheme for the estimation of the pressure gradient of the gas and the system self-gravity. The gas exchanges the Newtonian force with a set of bodies which represent stars or planets. In an \sph{} algorithm the gas is sampled by means of a set of particles each containing local physical properties such as temperature, density, pressure gradient, and Newtonian force. These local properties are treated as the same in a local sphere, the \textit{smoothing length}. The pressure gradient and other key quantities useful to calculate the acceleration and the time variation of temperature are estimated by means of suitable interpolations over a neighborhood ensemble of points. To model an accretion disk around a star, we use the standard $\alpha$-model \citep{shakura1973} for the turbulent viscosity of the disk, as illustrated in the bottom table of Table~\ref{Table1}. The turbulent viscosity represents an approximate physical scheme which mimics the dynamics of the turbulence motions inside the gas, which, as a net result, leads to an inward transport of matter. This is achieved through the dissipation of the orbital motion of the gas \citep[e.g.,][]{papa2005,papa2007}.
In an $\alpha$-disk, the gas dynamics is influenced by an effective kinematic viscosity that can be expressed as
\begin{equation}
    \nu = \alpha_{SS} c_s H \ ,
\end{equation}
where $c_s$ is the local speed of sound and $H$ is the vertical scale parameter, i.e., the local vertical distance from the disk mid-plane where the density and pressure of the disk are significantly decreased. Therefore, $\alpha_{SS}$ represents a dimensionless parameter that characterizes the local strength of the disk viscosity. In our \sph{} code, we translate the kinematic viscosity in a classical SPH form where $\alpha_{SS}$ is related to the strength of an additional artificial pressure term in the gas Eulerian equations \citep[see][for details]{meru2012, picogna2013}. We refer also to \cite{rosotti2014} for several examples of numerical models of turbulent viscosity in protoplanetary disks around stars found in dense star clusters.

\subsection{Disk modeling in the star cluster} 

The initial setups are listed in the central table of Table~\ref{Table1}. We use different configurations to model the evolution of the disk. 
We refer to the models as M1, M2 and~M3. These refer to three types of encounters in the star cluster. M1 is a very close encounter (with a minimum distance between the encountering and host star $<1000$~AU), M2 is an intermediate-distance encounter (within a distance $<10,000$~AU and $>1000$~AU) and finally M3, a large distance encounter ($>10,000$~AU and $<40,000$~AU). These encounter distances give a general idea of how an encounter affects the host star and its circumstellar material. A general description of how different encounter distances affect  a system is provided in Section~\ref{sec:kpar}, and a more extensive discussion can be found in \cite{flaetal2019} and \cite{spurzem2009}.\\
The next set of sub-models refers to the rotation and counter-rotation of the protoplanetary disk. The dynamical evolution of the star cluster, and the consequent close encounters, may change the evolution of the disk. Moreover, according to \cite{sanchez2018},  the radial migration timescale of inner objects in the retro-grade rotation may be appreciably shorter than in the pro-grade rotation. We will verify this statement in our work.\\
Finally, we will have three final sub-models, where we will add a Neptune-mass planet, at three different semi-major axes. These values are 30~AU (similar to Neptune in our own solar system), 50~AU (at the disk's half-mass radius) and 70~AU (in the outskirts of the denser section of the disk).
The addition of a planet is fundamental to answer two main questions: (i) whether the dynamical evolution of the planet is changed in respect of the absence of both star cluster and protoplanetary disk and (ii) whether the final outcome after the encounter depends on the initial conditions.

\section{Results}\label{results}

In this section we describe the evolution of the star cluster. After that, we will analyse the gas ejected from the disk and accreted by the stars, in order to evaluate its role in the orbital evolution of the planet. We will classify the star cluster's encounters in three different classes, based on the predicted impact on the protoplanetary system, which will be indicated by the $k$-parameter \citep{spurzem2009, flaetal2019}, described also in Section~\ref{sec:kpar}. Finally, we study the dependence on the properties of the protoplanetary disk system.

\subsection{Star cluster evolution}

\begin{figure}
    \includegraphics[width=0.5\textwidth,height=!]{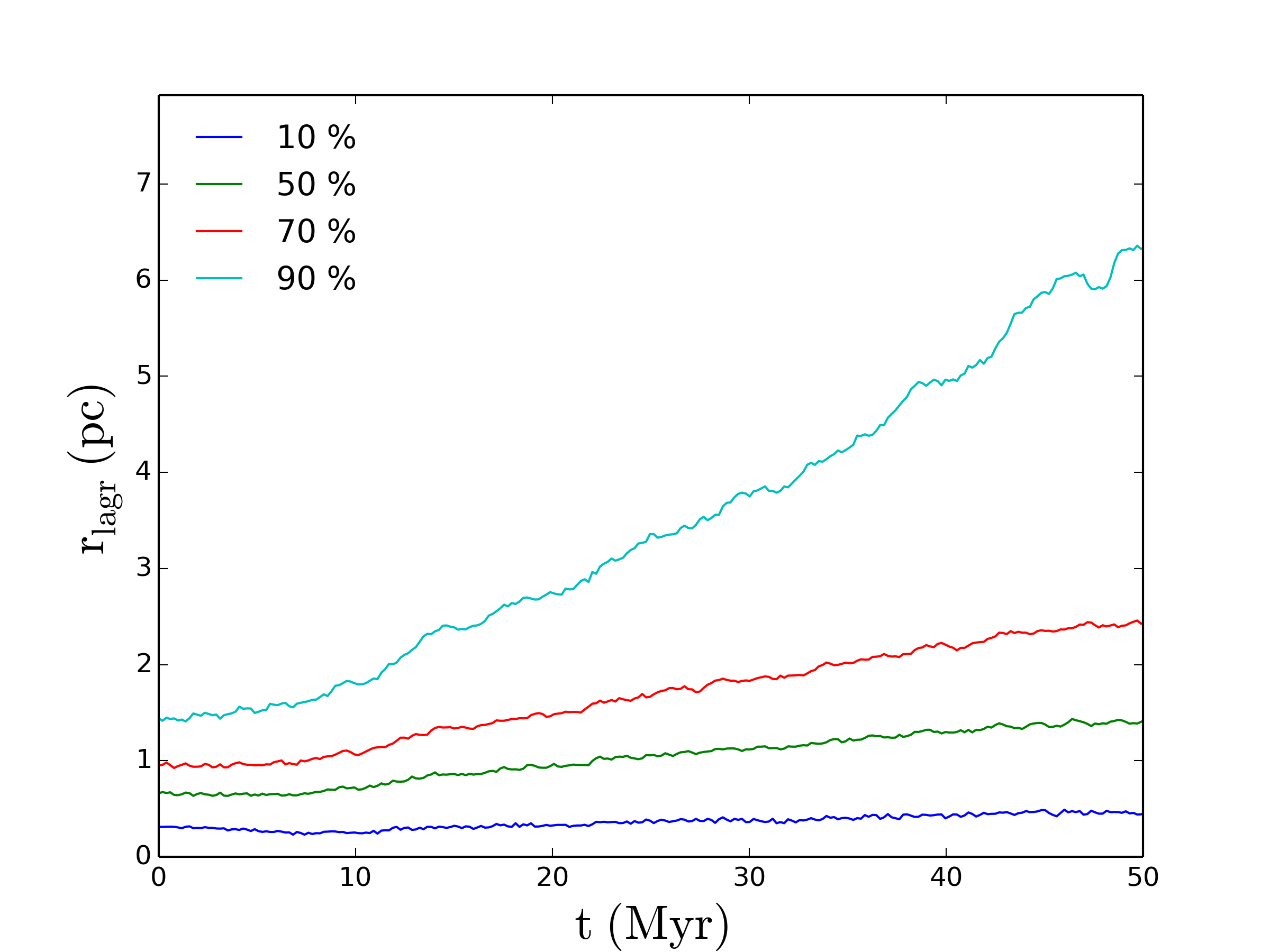} \\
    \caption{Evolution of the Lagrangian radii of the star cluster.}
    \label{Figure2}
\end{figure}

The evolution of the Lagrangian radii of the star cluster is shown in Figure~\ref{Figure2}. Any star cluster substructure is removed on the order of several crossing times, and the cluster enters into virial equilibrium on these timescales \citep[e.g.,][]{allison2009}. The clusters starts to expand around an initial half-mass relaxation time, resulting in the the ejection of cluster members. At the same timescale, the more massive stars start to sink towards the star cluster centre, and the low-mass stars start to migrate to the outskirts \citep{khalisi2007}. These processes are visible in the 90\% and 70\% Lagrangian radii. The 90\% shell grows more abruptly due to the star cluster filling its Roche lobe in about 10~Myr from the start of the simulation.

\subsection{Close encounters} \label{sec:kpar}

\begin{figure}
    \includegraphics[width=0.5\textwidth,height=!]{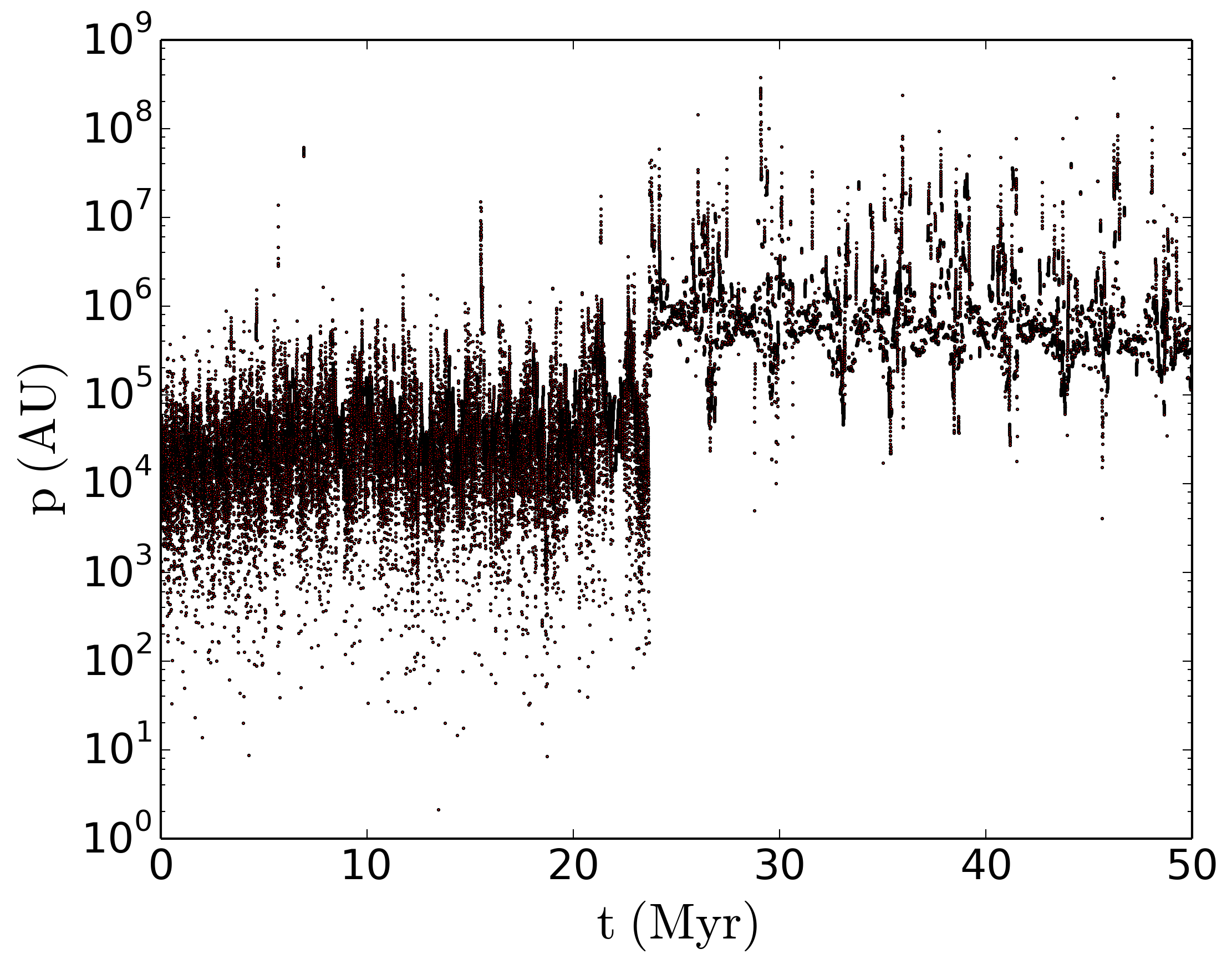} \\
    \caption{Temporal distribution of the instantaneous periastron distance $p$ for stellar encounters with the nearest neighbour experienced in the star cluster model C05.}
    \label{Figure3}
\end{figure}

\begin{figure}
    \includegraphics[width=0.5\textwidth,height=!]{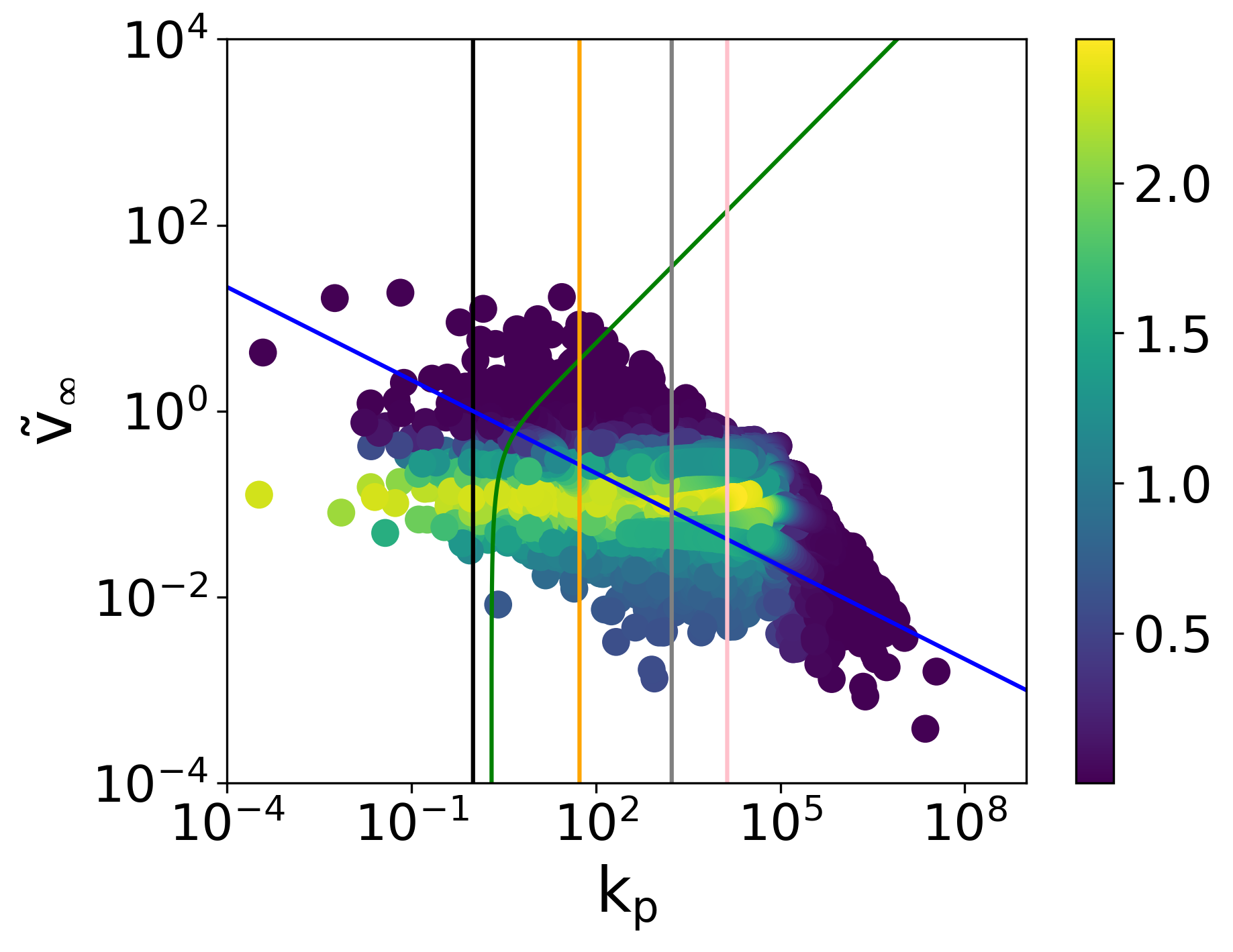} \\
    \caption{The distributions of the encounter strength parameter $k$ of the nearest stellar encounters, experienced by disk in model~C05, at the disk's half-mass radius, $\sim 50$~AU. The black curve separates tidal (right) and impulsive (left) encounters, the blue curve separates hyperbolic (above the curve) and near-parabolic encounters (below the curve), while the green curve separates adiabatic (right) and non-adiabatic (left) encounters. The curves are obtained from the properties of individual stellar encounters. The different data-point colours indicate the distribution density, with the most frequent encounters indicated in red, while less frequent encounters are indicated in blue. The ocher, grey and pink vertical curves are the distances of the encounter models chosen in this work (1\,000~AU, 10\,000~AU and 40\,000~AU) at their respective $k_p$ value. The color bar ranges from 0.0001 to 2.5. } 
    \label{Figure4}
\end{figure}

The main star cluster properties that characterize the evolution of the architecture of a planetary system in a star cluster are the star cluster density, the encounter strengths and the relative velocities of the encountering stars. From the perspective of planetary systems in a dense stellar environment, the semi-major axis is the most important property that determines its dynamical evolution under the influence of stellar encounters.\\
In our study we model protoplanetary disks with a Neptune-mass planet at three different semi-major axis: 70~AU, 50~AU and 30~AU. In the encounter analysis below, we consider a test particle at a semi-major axis of 50~AU. The properties of the stellar encounters are obtained from the star cluster model~C05.\\
Before analysing the encounter strength parameter $k$ \citep{flaetal2019, spurzem2009}, we analyse the evolution of the periastron distances in Figure~\ref{Figure3}. 
The color bar indicates the Gaussian kernel-density estimation of a dataset (KDE) in a bi-dimensional space ($k_p$,$\tilde{v_{\infty}}$) normalised by $10^{-5}$ for cosmetic purposes. This value is often used in statistical science to compare the the weight of more-than-one dimension parameters. Such points are weighted according to the presence of points in the parameter space. Larger values indicates more datapoints (the value of 2.5 is approximately 3000 datapoints, while the minimum is less than 1 datapoint). 
The periastron evolution suggests that encounters at nearby distances, until a $\sim t_{\rm rh}$, are more frequent compared to later times, where the probability of a close encounters drastically drops. The typical periastron distances of encountering stars are $\sim 10^{4}$~AU before a relaxation time, while they are $\sim 10^{6}$~AU after one relaxation time has passed. Close encounters with periastron distances below $1000$~AU occasionally occur. The reference model has, therefore, two different encountering distance phases, where nearby, and possibly more effective, close encounter distances may be detected. \\
Before analysing Figure~\ref{Figure4}, we first describe the quantities in the abscissa and ordinate. The quantity $\tilde{v_{\infty}}$ is defined as
\begin{equation}
    \tilde{v}_{\infty} = v_{\infty} \left( \frac{G(m_{hp}+m_n)}{a_p} \right)^{-1/2} \ .
    \label{eq:vinftilde}
\end{equation}
where $v_{\infty}$ is the velocity-at-infinity, $G$ is the gravitational constant, $m_{hp}$ is the host star mass, $m_n$ is the encountering star mass, and $a_p$ is the semi-major axis of the test particle. Equation~(\ref{eq:vinftilde}) quantifies the ratio between the velocity of the neighbour star and the planet orbital velocity. The parameter $k_p$ is defined as
\begin{equation}
    k_p = \sqrt{\frac{2 m_{hp}}{m_{hp}+m_n} \left(\frac{p}{a_p}\right)^3  } 
    \approx  \left(\frac{p}{a_p}\right)^{3/2} 
    \  .
    \label{eq:kparameter}
\end{equation}
where $p$ is the periastron velocity of the neighbour star.
If $m_{\rm hp} \approx m_n$, then the mass factor in Equation~(\ref{eq:kparameter}) can be considered negligible.\\
Figure~\ref{Figure4} shows the classification of the encounter strength using the $k$ parameter plotted against the normalised velocity-at-infinity. Most encounters are adiabatic, tidal and parabolic. The test particle is located at a semi-major axis of 50~AU, the corresponding distance of the half-mass radius of the protoplanetary disk. \\
The black curve in Figure~\ref{Figure4} separates impulsive ($p/a<1$) and tidal ($p/a>1$) encounters. The ratio between the periastron and semi-major axis of the test particle reflects how near the encountering star and test particle effectively are. This property is the most influential, as these kind of encounters have a larger probability of disrupting planetary systems.\\
The green curve in Figure~\ref{Figure4} separates adiabatic (right) and non-adiabatic (left) encounters. The ratio between $v_{\infty}$ and the orbital velocity of the test particle $\sqrt{G(m_{hp}+m_n)/a_p}$ is the key factor. If the velocities are comparable or below this value, the encounter is non-adiabatic, i.e. the test particle is affected by the tidal (or physical) encounter. If they are not comparable, the encounter is adiabatic, i.e., the test particle is not affected by the encountering star. \\
The blue curve in Figure~\ref{Figure4} separates hyperbolic (above the curve) and near-parabolic (below the curve) encounters. Near-parabolic encounter are most likely to effectively affect the test particle, as their encounter time is larger than parabolic orbits.\\
The coloured curves Figure~\ref{Figure4} represent the selected distances for our disk reference models M1, M2 and~M3, respectively at distances $<1\,000$~AU (ocher), $<10\,000$~AU and $>1\,000$~AU (grey), and $>10\,000$~AU and $<40\,000$~AU (pink). In this plot we define the distances and the periastrons as equal.\\
We retrieve the cases shown in the next two section of the paper from (i) the most probable and effective encounters for the M1 class (red to black data-points and the points which are near-parabolic, non-adiabatic and impulsive, respectively) and (ii) we take the most effective encounters on that interval for the M2 and M3 classes.
Although results are shown for a particle with a semi-major axis of $50$~AU, the results for similar semi-major axes are comparable. A relatively large fraction of the encounters is impulsive. This should not come as a surprise, as particles with large semi-major axes are more likely to be perturbed and ejected, therefore impulsive encounters are common, even in such clusters \citep{flaetal2019}. 
If we vary the semi-major axis, then the points will shift to the right in the plot for lower semi-major axes, and to the left for higher semi-major axes. The main consequence is that the number of impulsive encounters increase and it directly proportional to the semi-major axis as $\propto a_p^{-3/2}$, assuming the same dynamical properties of the stars. \\
The dynamics of circumstellar disks is different from that of a planet. The disk density and the presence of other objects (such as planets) may alter the probability of an effective encounter on the protoplanetary disk evolution. In the following, we will test if a certain number of parameters may change the final outcome of an encounter, with the given initial conditions.\\

\subsection{M1 cases} \label{M1section}

\begin{table}
    \caption{Different initial conditions for the protoplanetary disk and planet models in the M1 encounter class: the model ID (column~1, using the syntax D-i, where $i$ the model number) is used for the majority of models except for $nd$ = no disk, the disk rotation direction, co-rotating ($+$) or counter-rotating ($-$) (column~2), the initial inclination of the disk (column~3), the initial phase of the planet (column~4), the initial semi-major axis of the planet (column~5). }
    \begin{tabular} {lcccccc}
    \hline
    Event ID & $r$ or $c$ & disk  & planet & planet \\
     & + / -  & inclination ($^\circ$) & phase ($^\circ$)   &   $a$ (AU) \\
    \hline
    Dnd1   & +   & 0  & --- & ---  \\
    Dnd2  & $-$  & 0  & --- & ---   \\
    Dnd3  & + & 90  & ---  & --- \\
    D1 & +   & 0 & 0  & 30  \\
    D2 & $-$   & 0 & 0   & 30  \\
    D3 & +   & 0 & 90   & 30 \\
    D4 & +   & 0 & 180  & 30  \\
    D5 & +   & 0 & 0   & 50   \\
    D6 & $-$   & 0 & 0   & 50   \\
    D7 & +   & 0 & 90   & 50    \\
    D8 & +   & 0 & 180   & 50       \\
    D9 & +   & 0 & 0    & 70    \\
    D10 & $-$   & 0 &  0 & 70   \\
    D11 & +  & 0 & 90    & 70    \\
    D12 & + & 0 & 180  & 70  \\
    D13 & +   & 90 & 0   & 70  \\
    \hline
    \end{tabular}
    \label{Table2}
\end{table}

\subsubsection{Models of protoplanetary disks and planets}

We explore the consequences of a single close encounter between two stars, for which we found the nearest approach at $\approx 20$~AU. We will refer to the two stars in the simulation as the host star (the star which hosts the protoplanetary disk and the planet) and the bullet star (the star encountering with the host star in the barycentric reference system), respectively. We carry out several simulations for each event, with different initial configurations for the host star system. We vary (i) the inclination of the disk and the orbit of the planet, with respect to the star orbital plane, (ii) the orientation of the disk and planet's orbit, (iii) the planet's semi-major axis. We list our models in Table~\ref{Table2}.\\
The models analysed in this section are strictly limited to the subset of stellar encounters with a periastron distance $\ll 1000$~AU. This class of encounters strongly affects the evolution of both the disk and the planet, since they tend to be {\it impulsive}. Encounters founded at larger distances have been widely studied \citep[e.g.,][and references therein]{spurzem2009, flaetal2019}. Moreover, the kind of encounter we look for is more common in large density environment, as young massive clusters and globular clusters. An overview on how different encounter distances affect the planetary systems has been provided in Section~\ref{sec:kpar}. We describe, in the context of the reference event, a space of parameters that leads a diverse final outcome for the planet after an encounter. A variation in the dynamical evolution of both the planet and disk is expected, due to the differences in the initial conditions. The parameter's variations we present in our models are described in as follows.
\begin{enumerate}
    \item Initial orbital phase of the planet: the planet is in a different initial position than the default position (which we define at $\varphi = 0^\circ$). For an event with similar initial conditions, a different initial position for the planet may result in a completely different outcome of the encounter event. 
    \item Spatial orientation of the planetary orbit and of the disk: we simulate both co-rotating and counter-rotating systems.
    The rotation direction of the system, with respect to the direction from which the encountering star approaches, may determine the dynamical fate of the planet. Moreover, the dynamical evolution of the disk is affected by this relative orientation: a larger fraction of gas is expected to be perturbed by the bullet star when it scatters with the disk 
    in a counter-rotating direction. 
    \item Inclination of the disk and planet's orbit in the reference system: the orientation of the system with respect to the direction from which the encountering star approaches the host star, affects how much of the circumstellar gas is scattered by the encountering star. Therefore, we may expect different consequences for different outcomes, depending on the mass and kinematics of the bullet star.
    \item Planet semi-major axes: we use three different models: \\
    (a) $a=30$~AU, similar to the semi-major axis of Neptune in our solar system; \\ 
    (b) $a=50$~AU, placed at the disk's initial half-mass radius; \\
    (c) $a=70$~AU, placed in the outskirts, beyond the denser section of the disk, which has its highest density at $\approx 50$~AU. \\
    Since both the disk and the planet are initialised with Keplerian orbits, the orbital speed of the planet decreases with semi-major axis. 
\end{enumerate}

\subsubsection{M1 models analysis}\label{M1sec}

\begin{figure}
    \includegraphics[width=0.49\textwidth,height=!]{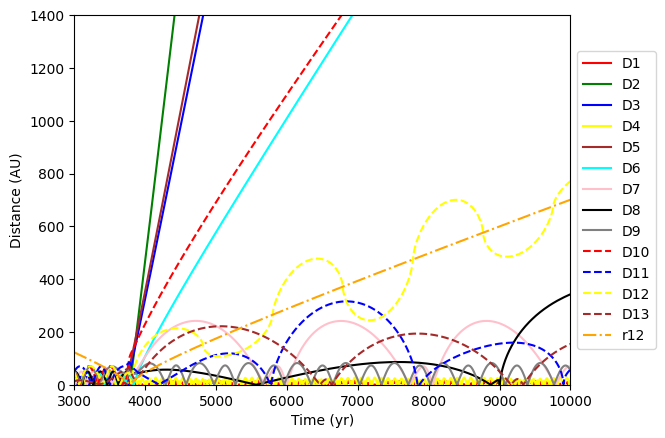} \\
    \caption[protoplanetary disks in star clusters: host star-planet distance]{Distance between the host star and the planet over timer all models that include a planet. The curve with the label r12 shows the distance between the host star and the bullet star. Except where mentioned otherwise, all models have a relatively similar semi-major axis compared to their initial conditions.}
    \label{Figure5}
\end{figure}

\begin{figure}
    \includegraphics[width=0.49\textwidth,height=!]{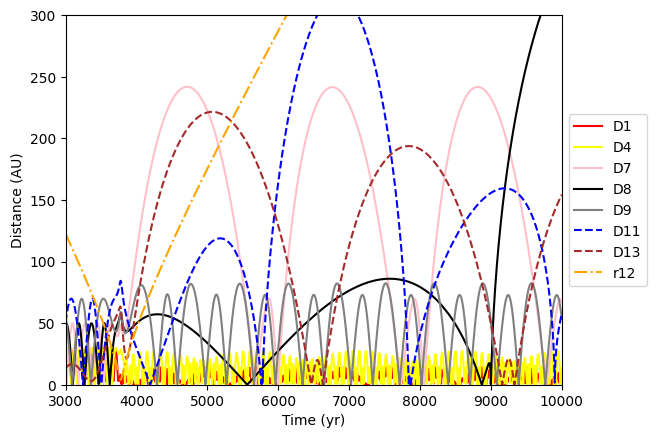} \\
    \caption[protoplanetary disks in star clusters: host star-planet distance]{Same as Figure~\protect\ref{Figure5}, zooming in the smaller distances from the host star. It is noticeable how the encounter history produced a different secular evolution for the bounded planets.}
    \label{Figure6}
\end{figure}

\begin{figure}
    \includegraphics[width=0.49\textwidth,height=!]{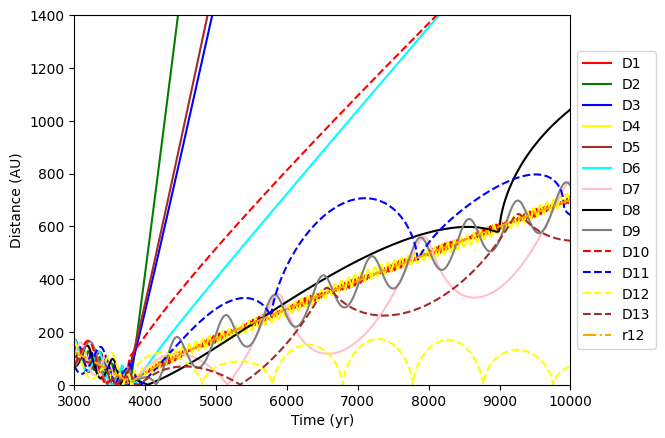} \\
    \caption[protoplanetary disks in star clusters:: encountering star-planet distance]{Distance between the bullet star and the planet over time, for all planet-including models. The curve labeled r12 shows the distance between the host star and the bullet star.}
    \label{Figure7}
\end{figure}

\begin{figure}
    \includegraphics[width=0.49\textwidth,height=!]{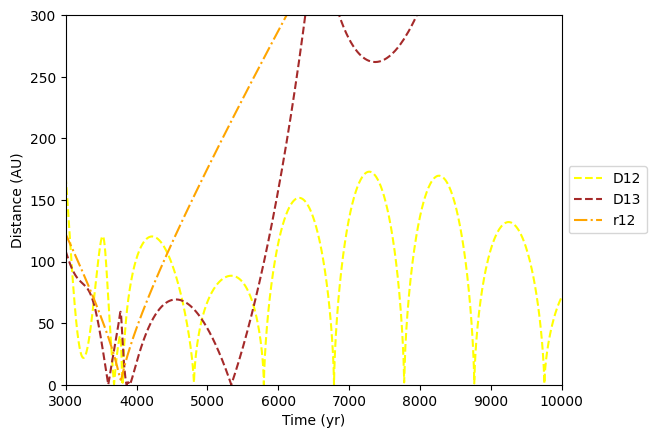} \\
    \caption[protoplanetary disks in star clusters: encountering star-planet distance]{Same as Figure \protect\ref{Figure7}, zooming in the smaller distances from the encountering star. Although at larger distances, model~D12 stays bound to the encountering star after being captured. It is quite noticeable, instead, model~D13, which has a larger orbit and goes nearby the encountering star, but it is still gravitationally bound to the host star.}
    \label{Figure8}
\end{figure}

We present several results obtained from the dynamical evolution of the different models listed in Table~\ref{Table2} and described in the section above. We first explore the evolution of planet-less models to compare the gas loss evolution with similar models. Then we focus on the other models. The host star-planet distances are shown in Figures~\ref{Figure5} and~\ref{Figure6} and the bullet star-planet distances are shown in Figures~\ref{Figure7} and~\ref{Figure8}. Below, we discuss our main findings for each of these models.

\begin{figure}
    \includegraphics[width=0.49\textwidth,height=!]{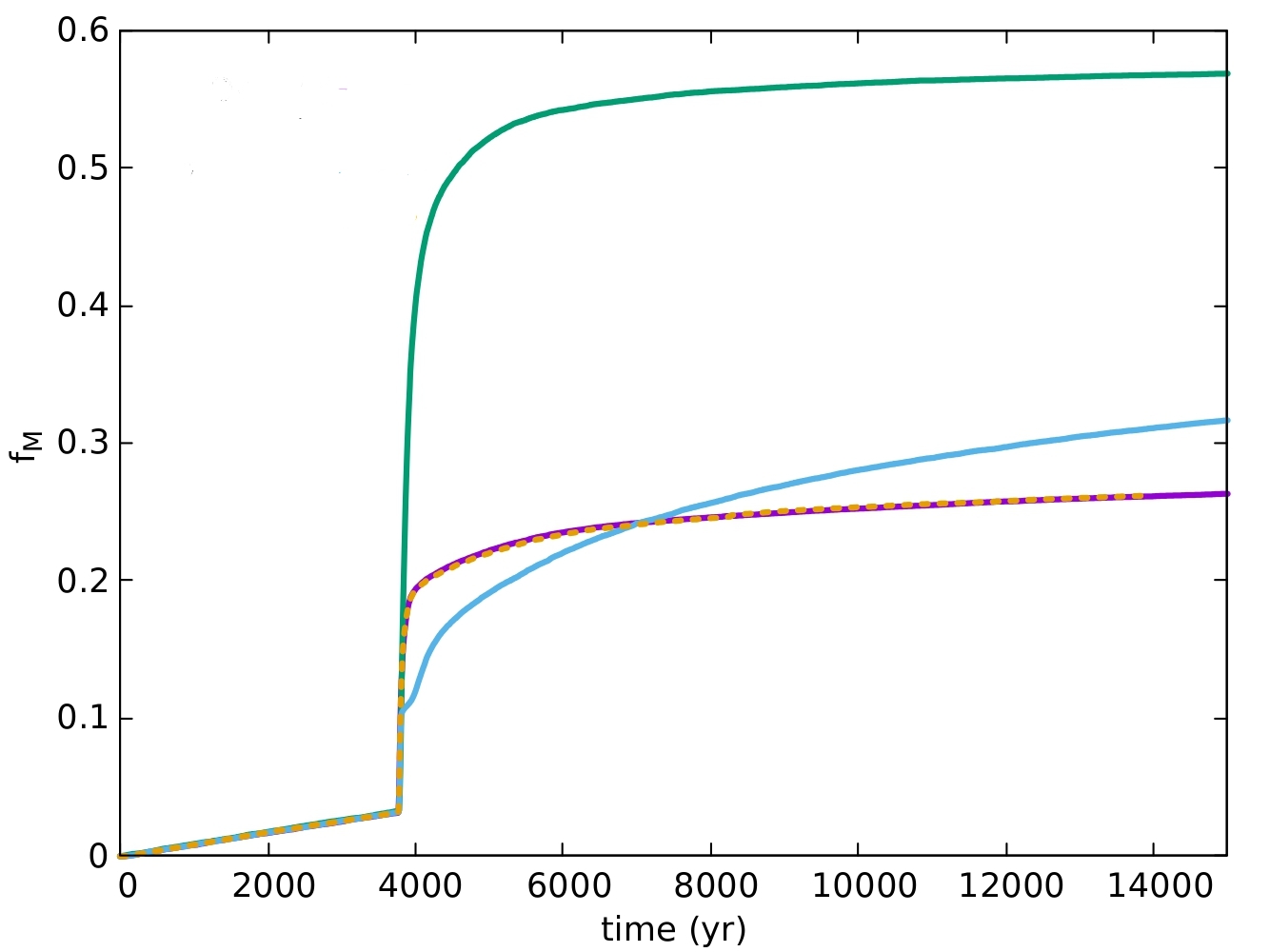} \\
    \caption[protoplanetary disks in star clusters: Gas captured by the host star]{Fraction of the disk mass captured by the host star, in models Dnd1 (purple), Dnd2 (green), Dnd2 (cyan) and~D1 (dotted orange). Model~D1 has a slighter higher gas capture fraction, which could have a prominent effect over long periods, but which is not relevant in the time frame of a single encountering event.}
    \label{Figure9}
\end{figure}

\begin{figure}
    \includegraphics[width=0.49\textwidth,height=!]{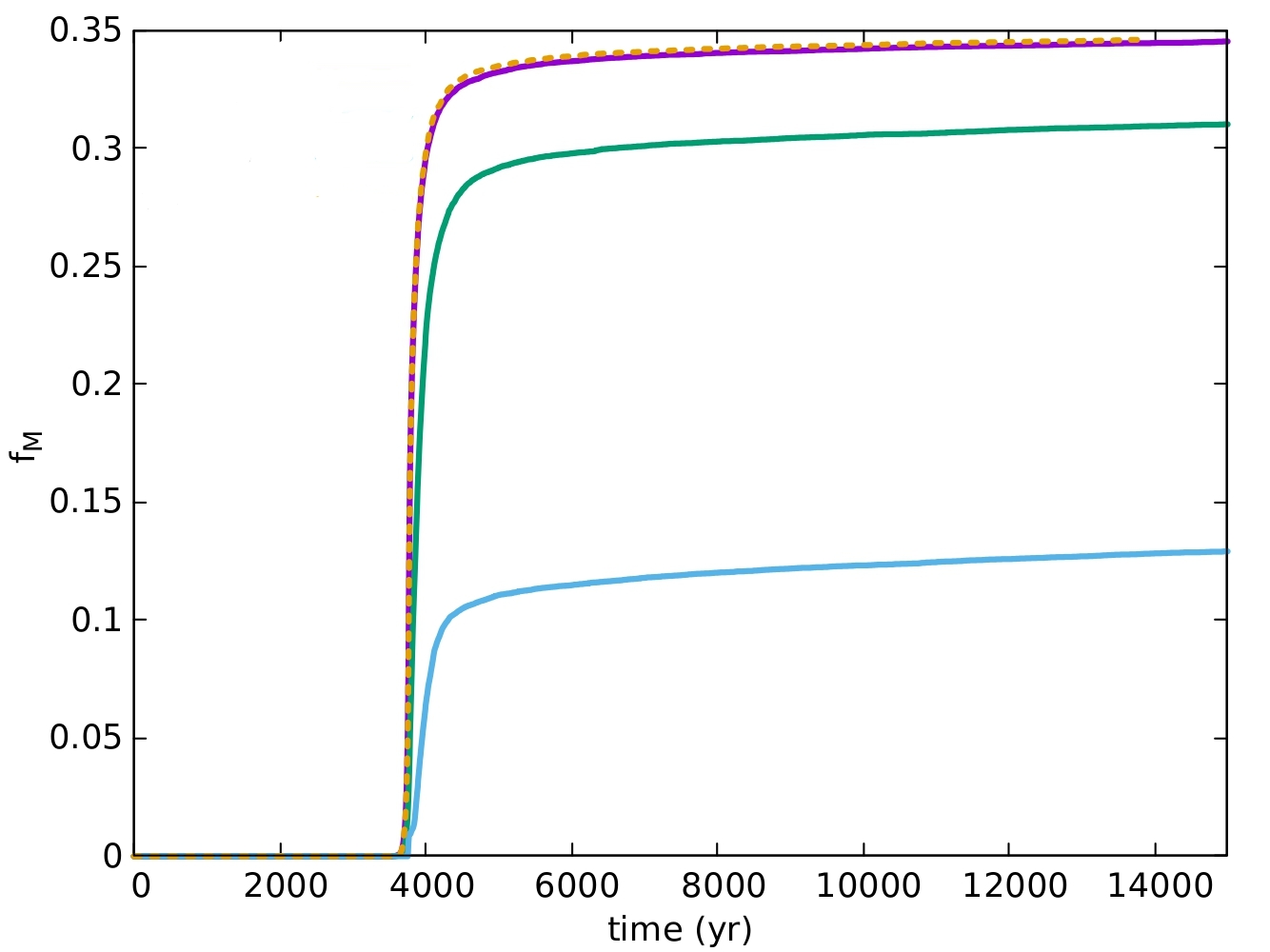} \\
    \caption[protoplanetary disks in star clusters: Gas captured by the bullet star]{Fraction of disk mass captured by the encountering star, in models Dnd1 (purple), Dnd2 (green), Dnd2 (cyan) and~D1 (dotted orange). Similarly to Figure~\protect\ref{Figure9}, the presence of the planet slightly enhances the fraction of gas captured by the bullet star.}
    \label{Figure10}
\end{figure}

\paragraph{Planet-less models}
Models Dnd1, Dnd2 and Dnd3 are modeled without a planet. The disks in these models are oriented in a co-rotating, counter-rotating, and perpendicular plane with respect to the encountering star, respectively. Figure~\ref{Figure9} shows that the quantity of gas captured by the host star in models Dnd3 and Dnd1 is smaller than in model Dnd2. This may be a consequence of gas particles that, in a counter-rotating disk, migrate inwards faster than in models with co-rotating disks \citep{breslau2019}. The difference between the quantity of gas migrating inwards for models Dnd3 and Dnd1 is likely a direct consequence of the different inclinations of the disk. This characteristic leads to less angular momentum exchange with the gas and a smaller scattering area for perturbation of the disk by the bullet star when the disk is perpendicular to the orbital plane. Therefore, we observe that, initially, the model Dnd1 absorbs more gas. However, the impact may have altered the orbits of the gas particles in the innermost regions, causing them to be accreted by the host star.
Figure~\ref{Figure10} is consistent with this hypothesis. The bullet star in model Dnd3 accretes a larger quantity of gas than the host star, but less than in the other two models, due to its smaller scattering area. In the counter-rotating and co-rotating models, the bullet star accretes a larger quantity of gas, as both of these models have co-planar disks. The host star in the co-rotating model accretes more gas than the host star in the counter-rotating model, due to the slightly longer duration of the fly-by, in the frame of the individual gas particles.

\paragraph{Different semi-major axis models}
The planets in models D1, D5 and~D9 have semi-major axes of 30, 50 and 70~AU, respectively. These models have co-planar and co-rotating disks, with a planet in a circular orbit. We will describe these models below.
  
    \begin{enumerate}
        \item D1: the planet semi-major axis is $30$~AU. After the encounter, the planet remains gravitationally bound to the host star, and obtains a highly eccentric orbit ($e \sim 0.99$). We compare the difference between the amounts of accreted gas by the host star and the bullet star in models without planets, and model~D1, in Figures~\ref{Figure9} and~\ref{Figure10}, respectively. Here, model Dnd1 can be compared to model~D1. The amount of accreted gas is similar in both models, with a slightly larger (0.5\%), percentage of gas absorbed by the bullet star. The gas absorbed by the planet in model~D1 is negligible.
        \item D5: the planet semi-major axis is $50$~AU. The planet is ejected with the highest velocity among all of the models we dynamically evolved. After the encounter, the planet's relative speed is $v_{\rm pl}\sim$6~km/s, with respect to both of the stars.
        The planet has a smaller initial orbital speed due to its larger semi-major axis. Therefore, the planet will have a different phase, as compared to the planet in model~D1. The orbital phase at the moment of the close encounter strongly influences the dynamical fate of the system.
        \item D9: the planet semi-major axis is $70$~AU. The planet appears to be weakly perturbed by the encounter, with a final eccentricity of $e_{\rm fin} \sim 0.09$.
        Small differences in the planet's orbital phase at the moment of the encounter, may result in significant differences in the planet's orbital parameters after the encounter. A large semi-major axis increases the probability of a close encounter with the bullet star. Additionally, the planet is less gravitationally bound to the central star, and therefore more easily affected by smaller changes in the gravitational field. 
    \end{enumerate}

\paragraph{Counter-rotating disk models}
Models D2, D6, D10 have the same initial conditions as models D1, D5 and~D9 respectively, but a counter-rotating disk and planet. In other words, the orbital phase is inverted due to the opposite direction of orbital motion. 
    The encounter causes an ejection of the planet in all models. Note that in this case the dynamical outcome completely changes, due to counter-rotation of the disk. The orbital eccentricity before the ejection grows for larger semi-major axes, suggesting that both quantities have an effect on the planet's dynamical fate.

\paragraph{Different orbital phases models}
Models~D3, D4, D7, D8, D11 and~D12 have the same initial conditions as the model D1 (for D3 and~D4), D5 (for D7 and D8) and D9 (for D11 and D12), but their planets have different initial orbital phases: 90$^\circ$ and 180$^\circ$, respectively. The planet is perturbed similarly in the different phases cases, with one exception in the smaller semi-major axis.
    The final eccentricity for the models with $a=70$~AU is $e_{\rm fin} \sim 0.97$ for model D11 and $e_{\rm fin} \sim 0.99$ for model D12. Similarly, the models with  with $a=50~$AU have $e_{\rm fin} \sim 0.60$ for model D7 and $e_{\rm fin} \sim 0.97$ for model D8. For models with $a=30$~AU, the model D3 has its planet ejected and model D4 has $e_{\rm fin} \sim 0.35$. Therefore, in the D11, D12, D7 and D8 models, the different orbital phase have similar consequences to the default position models D9 and D5 respectively. After the encounter, the planet in model D11 has a wider orbit (with an apoastron up to 215~AU) than the planet in model D12 (with an apoastron up to 87~AU). The planet in model D8 has a wider orbit (with an apoastron up to 300~AU) than the planet in model D7 (with an apoastron up to 161~AU). In both cases, the phase indicate a migration of the planet to more eccentric orbits and relatively large apoastron, as a consequence of such eccentricity. In models D3 and D4, we have an ejection and a relatively less perturbed planet in model~D4, with $e_{\rm fin} \sim 0.35$ resulting in an apoastron at 20~AU, which corresponds to internal migration. Therefore, the semi-major axis of the planet plays a key role in these models with different initial orbital phase. The dynamical evolution of the planet is drastically different for smaller semi-major axes, mostly near to the encounter periastron value ($\approx 20$~AU). We note the important of the orbital phase of the planet, which has a strong influence on the outcome of the dynamical interaction with the neighbouring star.
    
\paragraph{Perpendicular disk model}
Model~D13 has the same initial conditions as model~D9, but the disk is perpendicular to the equatorial plane. The encountering star  scatters the planet into a wider, and more eccentric, orbit. However, the planet remains gravitationally bound to the host star.  These models have their planet perturbed more {\it efficiently} than in model~D9, in which the planet obtains an eccentric orbit with $e_{\rm fin} \sim 0.76$.

\begin{figure}
    \includegraphics[width=0.49\textwidth,height=!]{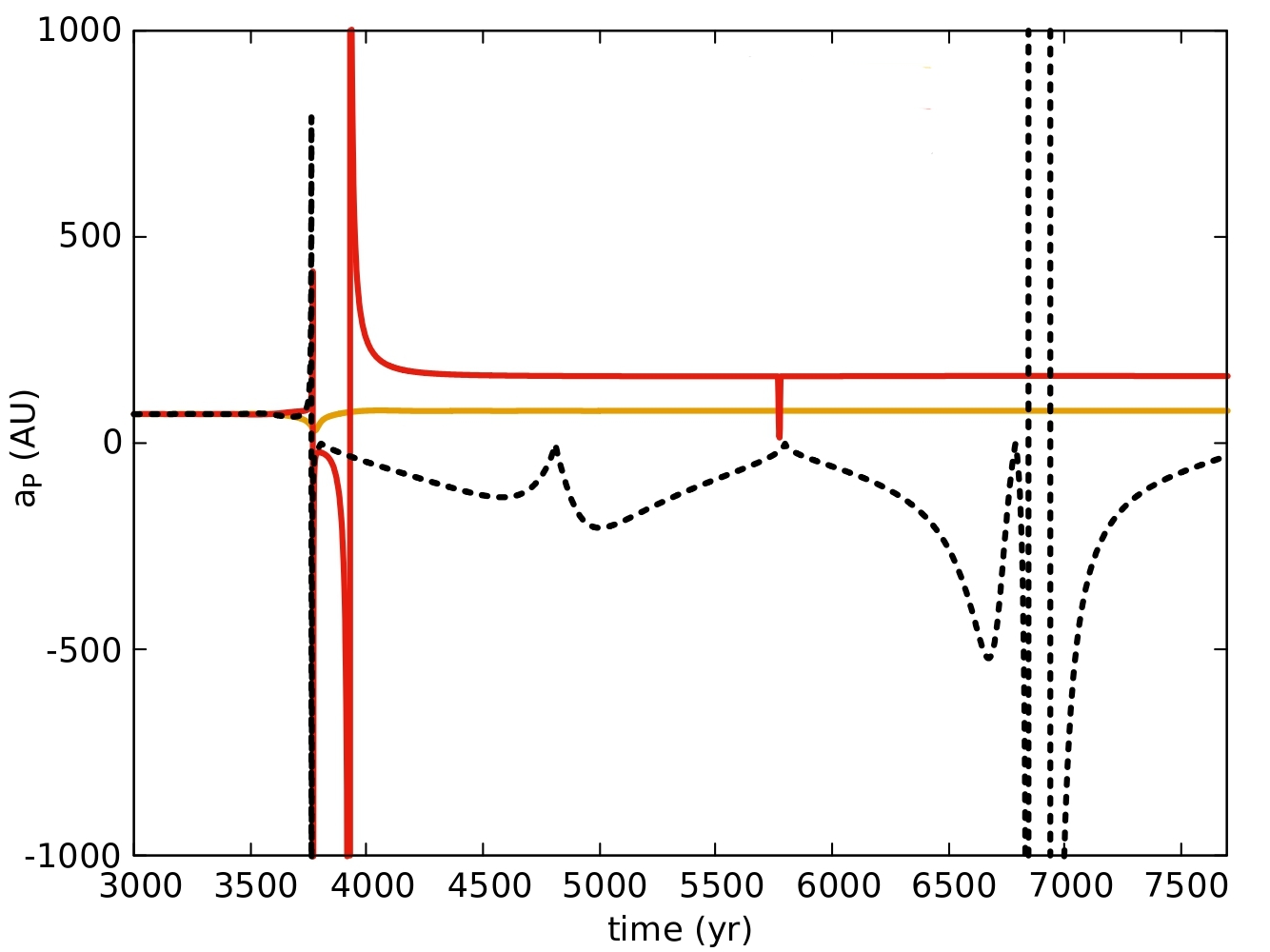} \\
    \caption[protoplanetary disks in star clusters:: different semi-major axis evolution for different phase models]{Models D9 (Orange), D11 (red), and~D12 (dotted black): evolution of the semi-major axis of the planet, before and after the encounter with the bullet star. A negative semi-major axis indicates that the planet is neither bound to the host star nor to the bullet star.}
    \label{Figure11}
\end{figure}

\subsection{M2 and M3 cases}

Unlike the M1 cases we have discussed in Section~\ref{M1sec}, the cases of M2 and M3 are related to the contribution of the far neighbour stars on the host star. The description of the disk models can be found in Table~\ref{Table3}. The main focus of this section is to identify the role of the star cluster environment on the evolution of the planetary orbit and the disk. \\
We choose just the larger semi-major axis between all the possible models due to it being the most subject to external perturbations. We will retain the same inclination for the planet and the disk models in these two cases, and we do not change either the eccentricity or the planet phase. The effect of more distant stars on these orbital parameters is negligible. Nevertheless, we explore the effect of counter-rotation for the nearest class of encounters of this section, M2. \\
In the cases of M2 and M3, there is no important difference in the short-term evolution, where the disk and planet tend to remain relatively unperturbed. The dynamical evolution of the systems differs from those of field stars only over longer periods of time. Compared to earlier works on the same topic \citep[e.g.,][]{rosotti2014}, we also analyse the effect on the planet. The results are comparable with \cite{hao2013}, and the planet is poorly affected in the long-term evolution, similarly to what was predicted in \cite{flaetal2019}. The D14 and D15 cases are mostly similar, except for the inverted rotation of the disk-planet system. The effect of the star cluster is weaker in D16.

\begin{table}
    \caption{Different initial conditions for the protoplanetary disk and planet models in the M2 and M3 encounter classes: the model ID (column~1, using the syntax D-i, where $i$ the model number), the disk rotation direction, co-rotating ($+$) or counter-rotating ($-$) (column~2), the initial inclination of the disk (column~3), the initial phase of the planet (column~4), the  initial eccentricity of the planet (columns~5), and the initial semi-major axis of the planet (column~6).}
    \begin{tabular} {lrccccc}
    \hline
    Event ID & $r$ or $c$ & disk &   planet  & planet \\
      & + / -  & inclination ($^\circ$) & phase ($^\circ$)    &   $a$ (AU) \\
    \hline
    D14 & +   & 0 & 0  & 70  \\
    D15 & $-$   & 0 & 0  & 70  \\
    D16 & +   & 0 & 0  & 70 \\
    \hline
    \end{tabular}
    \label{Table3}
\end{table}

\section{Discussion and conclusions}\label{conclusions}

Most stars form in dense stellar environments, where the gravitational influence of neighbouring stars may affect the planet formation process and the early evolution of planetary systems. In this study we analysed the effects of both close encounters and the long distance encounters on a protoplanetary disk containing an embedded planet.\\
Our goals were (i) to investigate the evolution of gas distribution of the disk due to the effects of the encounters and (ii) to study the consequences of both the stellar and disk perturbation on the planet orbit, as compared to a similar system in a pure $N$-body framework (i.e., a star without a protoplanetary disk). Our findings can be summarised as follows.

\begin{itemize}

    \item[$\circ$] The presence of a protoplanetary disk significantly impacts the dynamical fate of the planet. Our work suggests that the orbit of the planet is less perturbed by a close encounter when a protoplanetary disk is present. We will carry out a more comprehensive study on this matter in the near future to further prove this point.
    \item[$\circ$] Encounters at $r_p / a \leq \ 100$  are the ones which contribute more to modify the architecture of a protoplanetary system with a planet in a short time ($<1$~Myr). Distant encounters (i.e., $r_p / a \gg \ 100$ )
    are less important in the perturbation of the dynamical evolution of the planet and the disk. 
    
    \item[$\circ$] All the parameters we varied in the simulations have an impact on the dynamical fate of the planetary systems. The semi-major axis and the planet's initial orbital phase have the strongest impact. The influence of the inclination depends on the direction from which the encountering star approaches, and determines the effective scattering area.\\
    Below, we provide a description of the role of each parameter:
    
    \begin{enumerate}
    
        \item The difference between the orbital phases of the planets in the different models affects the dynamical outcome, after the interaction with the bullet star has occurred.
        A clear example is shown in Figure~\ref{Figure11}, which illustrates the evolution of the planet's semi-major axis in models~D9,  D11 and D12, respectively. In these models, the planets have an initial orbital phase of 0$^\circ$, 90$^\circ$ and 180$^\circ$, respectively. In model~D12, the planet is captured by the bullet star. In the other models, the semi-major axis increases significantly, although the planet is still gravitationally bound to its host star. It is clearly shown that the dynamical outcome after the encounter is radically changed when the initial orbital phase of the planet is changed. 
        \item The semi-major axis also plays an important role. The essentially Keplerian motion of the planet leads to different orbital velocities at different semi-major axis. Therefore, a different choice for the initial distance between the planet and the host star changes the configurations of the relative position and velocity of the planet with respect to the bullet star, during the encounter. The collisional cross section is also larger due to the larger semi-major axis.
        \item The inclination of the disk and the planet (relative to the orbital plane of the encountering star) appears, also, to be important. The inclination and the mass scattered away from the gaseous disk may be intrinsically related, and this represent an important difference on the final outcome of the encounter.
    \end{enumerate}
\end{itemize}

We observed both inward and outward migration, which will be studied in greater detail in future work. In this study, we have not taken into account the gravitational influence of the presence of multiple planets embedded in the circumstellar disk. We have also not taken into account the effect of radiation of the host star and of neighbouring stars on the disk. The presence of O/B stars in the star cluster substantially affect the evolution of the disk through photo-evaporation. Note that the latter is only important during the first few million years. Future studies are needed to extend the parameter space and to include a better treatment of distant neighbours in dense star clusters.

\section*{Acknowledgments}

We thank the referee for his invaluable help. 
FFD acknowledges support from the DFG priority program SPP 1992 “Exploring the Diversity of Extrasolar Planets” under project Sp 345/22-1. F.F.D. and M.B.N. were supported by the Research Development Fund (grant RDF-16--01--16) of Xi'an Jiaotong-Liverpool University (XJTLU). M.B.N.K. was supported by the National Natural Science Foundation of China (grant 11573004). We acknowledge the tremendous help of Luis Diego Pinto for the comments, suggestions and help with the use of the Gasph code.

\section*{Data availability}

The data underlying this article will be shared on reasonable request to the corresponding author.

\label{lastpage}
\end{document}